\documentclass[12pt,a4paper]{article}
\usepackage{amssymb,amsfonts,amsmath}

\title{Applying Quantum Principles to Psychology}

\author{Jerome R Busemeyer\\
Indiana University, USA\\
Zheng Wang\\
The Ohio State University, USA\\
 Andrei Khrennikov and Irina Basieva\\Linnaeus
University, Sweden}
\begin{document}

\maketitle

\begin{abstract}
This article starts out with a detailed example illustrating the utility
of applying quantum probability to psychology. Then it describes several
alternative mathematical methods for mapping fundamental quantum concepts
(such as state preparation, measurement, state evolution) to fundamental
psychological concepts (such as stimulus, response, information processing).
For state preparation, we consider both pure states and densities
with mixtures. For measurement, we consider projective measurements
and positive operator valued measurements. The advantages and disadvantages
of each method with respect to applications in psychology are discussed. 
\end{abstract}
\maketitle

\section{Why apply quantum theory to psychology?}

Twenty years ago, a group of physicists and psychologists introduced
the bold idea of applying the abstract principles from quantum theory
outside of physics to the field of human judgment and decision making
\cite{AertsAerts1994} \cite{Atmanspsacher2002} \cite{Bordely1998}
\cite{Khrennikov1999}. This new framework does \textit{not} rely
on the assumption that the brain is some type of quantum computer,
and instead it uses a probabilistic formulation borrowed from quantum
theory that involves non-commutative algebraic principles \cite{Khrennikov2010}
\cite{BusemeyerBruza2012} \cite{PothosBusemeyer2013}\cite{Wangetal2013}.
This new field, called \emph{quantum cognition}, has proved to be
able to account for puzzling behavioral phenomena that are found in
studies of a variety of human judgments and decisions including violations
of the ``rational'' principles of decision making \cite{PothosBusemeyer2009},
conjunction and disjunction probability judgment errors \cite{BusemeyerPothosFrancoTrueblood2011},
over- and under- extension errors in conceptual combinations \cite{Aerts2009}
\cite{GaboraAerts2002}, ambiguous concepts \cite{BlutnerBruzaPothos2013},
order effects on probabilistic inference \cite{TruebloodBusemeyer2011}
\cite{TruebloodBusemeyer2012}, interference of categorization on
decision making \cite{Busemeyer2009}, attitude question order effects
\cite{WangBusemeyer2013} and other puzzling results from decision
research \cite{Conte2009} \cite{LambertMogiliansky2009} \cite{LaMura2009}
\cite{YukalovSornette2011}. In short, quantum models of judgment
and decision have made impressive progress organizing and accounting
for a wide range of puzzling findings using a common set of principles.

\subsection{Example: Categorization-decision experiment}

To see more concretely how quantum theory can be applied to psychology,
consider the following psychology experiment used to investigate the
interference of categorization on decision making. Often decision
makers need to make categorizations before choosing an action. For
example, a military operator has to categorize an agent as an enemy
before attacking with a drone. How does this overt report of the category
affect the later decision? This paradigm was originally designed to
test a Markov model of decision making that is popular in psychology
\cite{Townsend2000}. Later it was adapted to investigate ``quantum
like'' interference effects in psychology \cite{Busemeyer2009}.

We begin by briefly summarizing the methods used in the experiments.
On each trial of several hundred training trials, the participant
is first shown a picture of a face that may belong to a ``good guy''
category (category G) or a ``bad guy'' category (category B), and
they have to decide whether to ``attack'' (action A) or ``withdraw''
(action W). The trial ends with feedback indicating the category and
appropriate action that was assigned to the face on that trial. There
are many different faces, and each face is probabilistically assigned
to a category, and the appropriate action is probabilistically dependent
on the category assignment. Some of the faces are usually assigned
to the ``good guy'' category, while other faces are usually assigned
to the ``bad guy'' category. The category is important because participants
are usually rewarded (win points worth money) for ``attacking''
faces assigned to ``bad guys'' and they are usually punished (lose
points worth money) for ``attacking'' faces assigned to the ``good
guys;'' likewise they are usually rewarded for ``withdrawing''
from ``good guys'' and punished for ``withdrawing'' from ``bad
guys.'' Participants are given ample training during which they learn
to first categorize a face and then decide an action, and feedback
is provided on both the category and the decision. Although the feedback
given at the end of each trial is probabilistic, the optimal decision
is to always ``attack'' when the face is usually assigned to a ``bad
guy'' category, and always ``withdraw'' when the face is usually
assigned to a ``good guy'' category. The key manipulation occurs
during a transfer test phase which includes the standard ``categorization
- decision'' (C-D) trials followed by either ``category alone''
(C-alone) trials or ``decision alone'' (D-alone) trials. For example,
on a ``decision alone'' trial, the person is shown a face, and simply
decides to ``attack'' or ``withdraw,'' and recieves feedback on
the decision. The categorization of the face on the D-alone trials
remains just as important to the decision as it is on C-D trials,
and some implicit inference about the category is necessary before
making the decision, but the person does not overtly report this implicit
inference.

Note that the C-D condition in the psychology experiment allows the
experimenter to observe which ``path'' the participant follows before
reaching a final decision. This is analogous to a ``double slit''
physics experiment in which the experimenter observes which ``path''
a particle follows before reaching a final detector. In contrast,
for the D-alone condition in the psychology experiment, the experimenter
does not observe which ``path'' the decision maker follows before
reaching a final decision. This is analogous to the ``double slit''
physics experiment in which the experimenter does not observe which
``path'' the particle follows before reaching a final detector.

According to the Markov model proposed in \cite{Townsend2000}, for
the D-alone condition, the person implicitly performs the same task
as explicitly required by the C-D condition. More specifically, for
the D-alone condition, once a face (denoted $f$) is presented, there
is a probability that the person implicitly categorizes the face as
a ``good'' or ``bad'' guy. From each category inference state,
there is a probability of transiting to the ``attack'' or ``withdraw''
decision state. So the probablity of ``attack'' in the D-alone condition
(denoted as $p(A|f)$) should equal the total probability of ``attacking''
in the C-D condition (denoted as $p_{T}(A|f)$). The latter is defined
by the probability that the person categorizes a face as a ``good
guy'' and then ``attacks'' plus the probability that the person
categorizes the face as a ``bad guy'' and then ``attacks'' ($p_{T}(A|f)=p(G\cap A|f)+p(B\cap A|f)$).
Using this categorization-decision paradigm, one can examine how the
overt report of the category interferes with the subsequent decision.
An\emph{ interference effect} of categorization on decision making
occurs when the probability of ``attacking'' for D-alone trials
differs from the total probability pooled across C-D trials. The Markov
model for this task originally investigated by \cite{Townsend2000}
predicts that there should be no interference, and the law of total
probability should be satisfied.

Beginning with our first study \cite{Busemeyer2009}, we have conducted
a series of four experiments on this paradigm. The results of these
experiments all generally show the same results, but we briefly report
a summary of findings from the fourth experiment that included 246
participants (a minimum 34 observations per person per condition).
When a face most likely is assigned to the ``god guy'' category
(we denote these faces as $g)$, the law of total probability is approximately
satisfied ($p_{T}(A|g)=.36,\: p(A|g)=.37$). However, when a face
most likely is assigned to the ``bad guy'' category (we denote these
faces as $b$), the probability of ``attack'' (i.e. the optimal
decision with respect to the average payoff) is systematically greater
for the D-alone condition as compared to the C-D condition'' violating
the law of total probability ($p(A|b)=.62>p_{T}(A|b)=.56$)%
\footnote{This difference are statistically significant: $t(245)=4.41, p=.0004$.
Also this same effect was replicated in 4 independent experiments%
}. More surprising, the probability of ``attack'' for the D-alone
condition (which leaves the ``good'' or ``bad'' guy category unresolved)
was even greater than the probability of ``attack'' given that the
person previously categorized the face as a ``bad guy'' ($p(A|b)=.62>p(A|b,B)=.61)$
on a C-D trial! For some reason, the overt categorization response
interfered with the decision by reducing the tendency to ``attack''
faces that most likely belonged to the ``bad guy'' category. These
violations of the law of total probability run counter to the predictions
of the Markov model proposed by \cite{Townsend2000} for this task.

\subsection{A Quantum decision model}

The details of a quantum model for the categorization-decision task
are presented in \cite{Busemeyer2009}, and here we only present a
brief summary. The human decision system is represented by a unit
length state vector $\left|\psi\right\rangle $ that lies within an
$4$-dimensional Hilbert space spanned by four basis vectors. Each
basis vector represents one of the four combinations of categories
and actions (e.g., $\left|GA\right\rangle $ is a basis vector corresponding
to category G and action A). The state $\left|\psi_{f}\right\rangle =\psi_{GA}\left|GA\right\rangle +\psi_{GW}\left|GW\right\rangle +\psi_{BA}\left|BA\right\rangle +\psi_{BW}\left|BW\right\rangle $
is prepared by the face stimulus $f$ that is presented on a trial.
The question about the category is represented by a pair of projectors
for good and bad categories $C_{G}=\left|GA\right\rangle \left\langle GA\right|+\left|GW\right\rangle \left\langle GW\right|,C_{B}=\left(I-C_{G}\right)$.
The question about the action is represented by a pair of projectors
for attack and withdraw actions $D_{A}=U_{DC}\left|GA\right\rangle \left\langle GA\right|U_{DC}^{\dagger}+U_{DC}\left|BA\right\rangle \left\langle BA\right|U_{DC}^{\dagger},D_{W}=\left(I-D_{A}\right)$,
where $U_{DC}$ is a unitary operator that changes the basis from
the categorization to the decision basis.

The probability of first categorizing the face as a ``bad guy''
and then ``attacking'' equals $p(B,A|f)=$ $p\left(B\right)\cdot p\left(A|B\right)$
= $\left\Vert C_{B}\left|\psi_{f}\right\rangle \right\Vert ^{2}\cdot\left\Vert D_{A}\left|\psi_{B}\right\rangle \right\Vert ^{2}$,
with $\left|\psi_{B}\right\rangle =\frac{C_{B}\left|\psi_{f}\right\rangle }{\left\Vert C_{B}\left|\psi_{f}\right\rangle \right\Vert }$,
and combining the terms in the product we obtain $p(B,A|f)=\left\Vert D_{A}\cdot C_{B}\cdot\left|\psi_{f}\right\rangle \right\Vert ^{2}$;
similarly, the probability of first categorizing the face as a ``good
guy'' and then ``attacking'' equals $p(G,A|f)=\left\Vert D_{A}\cdot C_{G}\cdot\left|\psi_{f}\right\rangle \right\Vert ^{2}$;
and so the total probability of attacking under the C-D condition
equals $p_{T}(A|f)=$ $\left\Vert D_{A}\cdot C_{G}\cdot\left|\psi_{f}\right\rangle \right\Vert ^{2}$
+ $\left\Vert D_{A}\cdot C_{B}\cdot\left|\psi_{f}\right\rangle \right\Vert ^{2}$.
The probability of attack in the D-alone condition equals $p(A|f)=\left\Vert D_{A}\cdot\left|\psi_{f}\right\rangle \right\Vert ^{2}=$
$\left\Vert D_{A}\cdot\left(C_{G}+C_{B}\right)\left|\psi_{f}\right\rangle \right\Vert ^{2}=$
$\left\Vert D_{A}\cdot C_{G}\left|\psi_{f}\right\rangle +D_{A}\cdot C_{B}\left|\psi_{f}\right\rangle \right\Vert ^{2}=$
$\left\Vert D_{A}\cdot C_{G}\left|\psi_{f}\right\rangle \right\Vert ^{2}$
+ $\left\Vert D_{A}\cdot C_{B}\left|\psi_{f}\right\rangle \right\Vert ^{2}$+
$Int$, where $Int=2\cdot Re\left[\left\langle \psi_{f}|C_{G}D_{A}C_{B}|\psi_{f}\right\rangle \right]$.
If the projectors for categorization commute with the projectors for
action (e.g., $U_{DC}=I)$, then the interference is zero, $Int=0$,
and we obtain $p(A|f)=\left\Vert D_{A}\cdot C_{G}\left|\psi_{f}\right\rangle \right\Vert ^{2}$
+ $\left\Vert D_{A}\cdot C_{B}\left|\psi_{f}\right\rangle \right\Vert ^{2}$
$=p_{T}\left(A|f\right),$ and the law of total probability is satisfied.
However, if the projectors do not commute (e.g., $U_{DC}\neq I$),
then we obtain an interference term. We can select the unitary operator
$U_{DC}$ to produce an inner product $Int=-.06$, and account for
the observed violation of the law of total probability.

We originally conducted these experiments because we predicted that
an interference effect of categorization on decisions would occur
based on past research using quantum models of decision \cite{BusemeyerWangTownsend2006}.
However, we could not predict the direction or quantitative size of
the interference. Now that we have this estimate, we can use it to
make new predictions for new experiments. Along this line, we carried
out a second condition in our fourth study to test our model. During
the transfer phase of the second condition, we included two different
types of transfer test trials: (a) D- alone trials as described earlier
in which the participant did not categorize but only made an action
decision, and (b) trials on which we informed the participant about
the correct category and the person made an action decision. Note
that for the second condition (b), the participants did not make any
categorization response and only made a decision, but this decision
was based on information provided by the experimenter about the category.
According to our theory, providing information about the category
produces the same effect as taking a measurement of the category --
if the person is told the face belongs to the ``bad guy category,''
then the state is updated from $\left|\psi_{f}\right\rangle $ to
$\left|\psi_{B}\right\rangle =\frac{C_{B}\left|\psi_{f}\right\rangle }{\left\Vert C_{B}\left|\psi_{f}\right\rangle \right\Vert }$
in the same way as if a measurement was made. Therefore, if we use
the choice probabilities from this second condition to compute the
total probability, then this second condition should produce exactly
the same interference effect as the first. Supporting this prediction,
the interference term for the second condition equaled $Int$= -.05,
which closely approximates the interference obtain from the first
condition. The great challenge for quantum cognition models is to
use the same principles then to predict new findings (e.g., see \cite{WangBusemeyer2013}\cite{WangInPress},
further a priori tests of the theory).

\section{Step by Step application of quantum theory to Psychology}

How does one apply quantum principles to psychological experiments?
How are the stimuli and responses of a psychology experiment related
to the state preparation and measurements of quantum theory? How does
one determine the observables, initial states, and unitary operators
for psychology? The above application involved several important assumptions
regarding the mapping of basic concepts in physics (e.g., state preparation,
state evolution, measurement operators) into basic concepts of psychology
(e.g., stimuli, information processing, responses). The next sections
examine these mappings more carefully, explores different ways to
formulate this mapping, and discusses the advantages and disadvantages
of different mappings.

\subsection{Choosing a Hilbert space}

Both physicists and psychologists are faced with the scientific task
of making predictions about the probability that different kinds of
events occur in their experiments. For example the psychologist wants
to predict whether a person will ``attack'' or ``withdraw'' when
a face is presented; a quantum physicist wants to predict whether
a particle is detected by one detector or another after it is emitted
from a source. Traditionally, psychologists have used classic theory
(axiomatized by Kolmogorov), whereas quantum physicists use quantum
theory (axiomatized by Dirac and von Neumann) to make these predictions.

Classic theory represents events as subsets of a universal set called
the sample space. Quantum theory represents events as subspaces of
universal vector space called the Hilbert space. Technically, a Hilbert
space is a vector space defined on a complex field and endowed with
an inner product that satisfies completeness. The adoption of a vector
space instead of a sample space to represent events may be the most
important assumption that is made in the application of quantum theory
to psychology.

Physicists work with both finite and infinite dimensional Hilbert
spaces. For example, an infinite dimensional space is used to represent
the position of a particle, but a finite n-dimensional space is used
to represent the spin of a particle. There is no a priori reason why
a psychologist could not use infinite dimensional spaces, but as in
the field of quantum computing, almost all of the previous work in
quantum cognition has used finite dimensional spaces. Choosing the
dimensionality of the Hilbert space is critical first step for constructing
a quantum cognition model. The dimension depends on the number of
variables and the number of values on each variable.

Two issues need to be considered when choosing the dimension of the
Hilbert space. The first issue is whether the measurements of a variable
are considered coarse (degenerate) or complete (non degenerate) (see,
e.g., \cite{Peres1998}). The second issue arises when there are two
or more variables and the question is whether or not the variables
should be combined to form what is called a tensor product space.

First consider the issue of completeness. The measurement of a variable
is complete if the outcomes cannot be refined, and so each outcome
can be represented by a single ray or a single dimension. Consider
an example such as asking a juror to rate the probability of guilt
on a five level scale (1 = very low probability, 2 = moderately low
probability, 3 = uncertain, 4 = moderately high probability, 5 = very
high probability). This variable is measured by five mutually exclusive
and exhaustive outcomes and so one might wish to represent this variable
by a 5 dimensional space. However, a person may be capable of rating
confidence on a much finer scale. Suppose the finest scale is a 21
level scale (0 = certainly not guilty, 10, ..., 50 = equally likely,
60, ..., 90, 100 = certain guilty). Then the 21 level scale forms
a complete measurement and requires a 21 dimensional space. If the
experimenter uses a 5 level scale, then this represents a coarse measurement
defined within a higher 21 dimensional space (e.g., levels 70 to 100
could be mapped into category 5). In reality, we do not know how fine
is the internal scale of the human decision maker. 

Next consider the tensor product space issue. Often in psychology
experiments, the participant is asked more than one question. Consider
our previous example of the categorization-decision experiment. In
this study, there are two variables: one is the categorization (categorize
as good versus bad) and the other is the action (choose between attack
versus withdraw). The problem is how to represent both variables within
a single vector space. The simplest possible representation is obtained
by using the same two dimensional space to represent both variables.
This is done by using a different basis to represent each variable
within the same two dimensional space. A more complex representation
is obtained by using a four dimensional space that represents each
combination of values for the two variables. The latter is called
a tensor product space.

How does one decide whether to represent the two variables within
the same space or to combine the variables into a tensor product state?
The tensor product representation assumes that it is possible for
a person to \emph{simultaneously} consider the values of \emph{both}
variables at the same time. The consideration of one value on one
variable does not disturb the evaluation of the second variable, and
there should be no order effects when the two variables are \emph{simultaneously}.%
\footnote{Simultaneity is important condition, because if a unitary transformation
occurs in between the measurements, then order effects can occur. %
} If this simultaneous evaluation is not possible, and the consideration
of the value of one variable disturbs the evaluation of the other
variable, producing order effects, then the variables have to be evaluated
sequentially by changing the basis within a common space.

Ultimately, picking the dimension becomes an empirical question. This
must be done by balancing accuracy and parsimony. One starts with
the simplest model, and if that fails empirically, then one is forced
to gradually increase complexity. In \cite{Busemeyer2009} we initially
tried the simplest 2 dimensional space with different bases for each
variable, but this failed empirically. Then we tried the next simplest
representation based a four dimensional (tensor product) vector space,
and this was empirically satisfactory.

\subsection{Choosing a basis to construct projectors and observables}

A basis is a set of orthormal basis vectors that span the Hilbert
space. Once we select a basis, we can construct a projector for the
subpace representing an event by using its basis vectors. The projector
for a subspace is formed by the sum of the outer products of basis
vectors that span the subspace. Consider once again the categorization-decision
task, and suppose that the four orthonormal vectors $\left(\left|GA\right\rangle ,\left|GW\right\rangle ,\left|BA\right\rangle ,\left|BW\right\rangle \right)$
form the basis for the categorization. These four vectors can be represented
numerically by $\left[1,0,0,0\right]',[0,1,0,0]',[0,0,1,0]',[0,0,0,1]',$
respectively. If we wish to measure the event ``bad guy category''
then we choose the basis vectors $\left|BA\right\rangle ,\left|BW\right\rangle $
representing this category, and form the projector $C_{B}=\left|BA\right\rangle \left\langle BA\right|+\left|BW\right\rangle \left\langle BW\right|$,
which numerically corresponds to the matrix $diag\left[0,0,1,1\right].$
Likewise we can define the projector for the ``good guy'' category
using the basis vectors $\left|GA\right\rangle ,\left|GW\right\rangle ,$
and numerically this projector corresponds to the matrix $diag[1,1,0,0]$.
If a set of projectors are pairwise orthogonal and sum to the identity,
then it forms a complete set that represents a mutually exclusive
and exhaustive set of events. In this example, the two projectors
are orthogonal, $C_{G}C_{B}=0$, and they sum to the identity, $C_{G}+C_{B}=I$,
and so they form a complete set.

Using the projectors of a complete set, we can form an observable
by assigning a real number to each projector. For example, suppose
we assign $+1$ to the ``bad guy'' event and $-1$ to the ``good
guy'' event; then the observable for the categorization variable
is defined as $C=(-1)\cdot C_{G}+(+1)\cdot C_{B}=C_{B}-C_{G}$. This
observable is constructed from the four categorization $\left(\left|GA\right\rangle ,\left|GW\right\rangle ,\left|BA\right\rangle ,\left|BW\right\rangle \right)$
basis vectors. These four basis vectors are eigenvectors of this observable
because each one satisfies the eigenvector equation, for example $C\left|BW\right\rangle =(+1)\cdot\left|BW\right\rangle $.
The same eigenvalue $(+1)$ is assigned to both eigenvectors $\left|BA\right\rangle ,\left|BW\right\rangle $.
A repreated eigenvalue like this is called a degenerate eigenvalue
-- any linear combination of eigenvectors $\left|BA\right\rangle ,\left|BW\right\rangle $,
which share the same $(+1)$ eigenvalue is also an eigenvector of
$C$ with the same $(+1)$ eigenvalue. A degenerate eigenvalue implies
that the observation of that value is a coarse measurement because
there is more than one basis vector associated with that value. 

There is, however, an infinite number of choices for the basis. Different
questions might require a different choice of basis to represent the
answers. Consider once again the categorization decision task described
earlier. One basis could be used to judge the strength of evidence
favoring each category (good versus bad categories), but a different
basis may be needed to evaluate the consequences of actions (attack
versus withdraw actions). In the above example, the unitary operator
$U_{DC}$ was used to change the basis from the one used for categorization
to the one used for action. The new basis for the action decision
is $\left(U_{DC}\left|GA\right\rangle ,U_{DC}\left|GW\right\rangle ,U_{DC}\left|BA\right\rangle ,U_{DC}\left|BW\right\rangle \right)$.
Then the projector for the ``attack action'' is $D_{A}=U_{DC}\left(\left|GA\right\rangle \left\langle GA\right|+\left|BA\right\rangle \left\langle BA\right|\right)U_{DC}^{\dagger}$.
Likewise the projector for the ``withdraw'' action is $D_{W}=\left(I-D_{A}\right)$.
If we assign $-1$ to the ``withdraw'' action and $+1$ to the ``attack''
action, then the observable is $D=D_{A}-D_{W}$. Note that the observable
for the action does not commute with the observable for the categorization
because the commutator $DC-CD\neq0$ does not equal zero.

The most difficult task is determing the form of the unitary operator,
$U_{DC},$ used to change the basis. A completely general form for
a unitary operator to a basis $Y=\left|Y_{i}\right\rangle ,i=1,N$
from another basis $X$ is $U_{YX}=\sum_{j}\left|Y_{j}\right\rangle \left\langle X_{j}\right|$,
but this does not give much guidance. One general way to construct
a unitary matrix is to use the exponential function of the operator
$H$, $U_{DC}=exp\left(-iH\right)$, where $H$ is a self adjoint
linear operator $\left(H=H^{\dagger}\right)$ called the Hamiltonian.
Any unitary operator can be formed in this manner. Using this method,
the problem then becomes one of choosing the form of the Hamiltonian.
In \cite{Busemeyer2009}, we designed a specific Hamiltonian based
on the rewards and punishments of the categorization - decision task.

If we present different faces but ask the same questions about categories
and actions for each face, then the same observables $C,D$ can be
applied to each face. Thus, according to this interpretation, the
observable only depends on the question and it does not depend on
the stimulus. We should point out, however, that the separation between
state preparation and selection of the observable is not always clear
cut, and one could argue that they can't be separated, so there are
alternative viewpoints on this issue that we discuss later.

\subsection{Preparing states}

In quantum theory, the system under investigation is represented as
a unit length vector, $\left|\psi\right\rangle $, in the Hilbert
space. In physics the system often refers to a particle, but in psychology
the system usually refers to a person. The state vector $\left|\psi\right\rangle $
can be expressed as a linear combination of the basis vectors and
the coordinates of the state vector with respect to a basis are called
the amplitudes. For example, referring back to the categorization
- decision task, if we choose the categorization basis, then the state
vector is defined by $\left|\psi\right\rangle =\psi_{GA}\left|GA\right\rangle +\psi_{GW}\left|GW\right\rangle +\psi_{BA}\left|BA\right\rangle +\psi_{BW}\left|BW\right\rangle $,
and $\psi_{GW}$ is for example the amplitude assigned to $\left|GW\right\rangle $.
Generally, these amplitudes can be complex numbers and the sum of
the squared magnitudes equals one. The interpretation of these complex
numbers is difficult for many psychologists, but there is no a priori
reason for limiting psychological applications to real numbers, just
as there is no reason to limit electrical engineering applications
to real numbers. In fact, Fourier analysis, using the complex transform,
is commonly used in both electrical control engineering, neural signal
processing, and human psychophysics. Utlimately the answers that we
obtain and need to interpet are always real.

In physics, the experiment begins with some physical system in some
state and then the experimenter prepares the state of the system by
applying physical devices before testing begins. Different types of
physical tests can be performed on systems after they are prepared
in the same state. In psychology, the person begins with some state,
and then the experimenter manipulates the state of the person by presenting
information or a stimulus prior to questioning. In our categorization
- decision experiment, the participant is presented a new face on
each trial, and the experimenter asks questions about the category
and the action to take for that face. Different types of questions
(which category, which action) can be asked about the same face. Therefore
the state before questioning is conditioned on the stimulus, $\left|\psi_{f}\right\rangle $,
where $f$ indicates that the person was shown a face labeled $f$
before asking any questions. Thus, according to this interpretation,
the presentation of a stimulus in a psychology experiment corresponds
to state preparation in a physics experiment and the state (before
the question) does not depend on the question that is asked later.
\emph{Any} question can be asked from this state, and the probability
of answers will vary across questions for the same state. Once again
we should point out that this separation between state preparation
and selection of the observable is debatable, and later we consider
are alternative viewpoints on this issue. 

There are two different ways to prepare the state in quantum physics.
One is by a measurement that projects the state to a subspace followed
by normalization to unit length. The other is by application of a
unitary operator that ``rotates'' the state in the Hilbert space
while maintaining unit length. Both of these methods could be used
in a psychology experiment.

The stimulus information that the participant experiences changes
the state of the person. For example, in the categorization-decision
experiment, the face that is presented at the beginning of a trial
will influence the state of the person before making a categorization.
For example, if the face looks like a ``bad guy'' the state will
move toward the subspace for that category. There are various ways
that this change could happen. One way is to use a unitary operator
$U_{f}$ that depends on the face stimulus. If the initial state,
before the face is presented, is define as $\left|\psi_{I}\right\rangle $,
then the state after presentation of the face becomes $\left|\psi_{f}\right\rangle =U_{f}\left|\psi_{I}\right\rangle $.%
\footnote{To be more precise, we should use the notation $\left|\psi_{I,f}\right\rangle =U_{f}\left|\psi_{I}\right\rangle $,
because the transformed state also depends on the initial state. But
to avoid using too many subscripts, and keeping in mind the history,
hereafter we use the shorter notation.%
}

The experimenter could present the participant some facts, which if
accepted to be true, would cause the person's state to be projected
onto the subspace consistent with those facts, and normalized to have
unit length. For example, in the categorization - decision experiment,
the participant could be informed that the face actually belongs to
the ``bad guy'' category before being asked to make an action decision.
Define $\left|\psi_{f}\right\rangle $ as the state after seeing the
face but before any information is presented on a trial. After the
new category information is presented, the state is updated to become
$\left|\psi_{B}\right\rangle =(1/c)\cdot C_{B}\left|\psi_{f}\right\rangle ,\: c=\left\Vert C_{B}\left|\psi_{f}\right\rangle \right\Vert $
and then the decision to attack is based on this updated state.

\subsection{Computing probabilities and updating states}

The purpose of using quantum theory for both physicists and psychologists
is to predict the probability of events. For a given state, the probability
of an event is obtained by projecting the state vector onto the subspace
for the event and computing is squared length. For example, referring
again to the categorization - decision study, if the person sees a
face and is asked to categorize it, then the probability of ``bad
guy'' equals $p(B|f)=\left\Vert C_{B}\left|\psi_{f}\right\rangle \right\Vert ^{2}$.
In physics, the state of the system changes following a measurement.
The same process occurs in psychology -- asking a question and deciding
on a definite answer changes the state of the person. According to
Lüder's rule, if the person categorizes the face as a ``bad guy,''
then the new state equals $\left|\psi_{B}\right\rangle =(1/c)\cdot C_{B}\left|\psi_{f}\right\rangle ,\: c=\left\Vert C_{B}\left|\psi_{f}\right\rangle \right\Vert $.
From this it follows that the probability of categorizing as a ``bad
guy'' and then deciding to ``attack'' equals $p(B,A|f)=$$\left\Vert D_{A}\cdot\left|\psi_{B}\right\rangle \right\Vert ^{2}\cdot\left\Vert C_{B}\cdot\left|\psi_{f}\right\rangle \right\Vert ^{2}=$
$\left\Vert D_{A}\cdot C_{B}\cdot\left|\psi_{f}\right\rangle \right\Vert ^{2}\cdot$
At the very end of the trial, with a categorization of ``bad guy''
and the decision to ``attack'' for example, the final state becomes
$\left|\psi_{AB}\right\rangle =\left(1/d\right)\cdot D_{A}\left|\psi_{B}\right\rangle $,
where $d=\left\Vert D_{A}\left|\psi_{B}\right\rangle \right\Vert $.

\subsection{Sequential effects}

An important question that still needs to be addressed concerns the
changes in the state of the person from one trial to the next. For
example, in the categorization - decision paradigm, each trial begins
with the presentation of a face and ends with feedback about the correct
category and action. This trial structure is part of the instructions
given to the participant at the beginning of the experiment. In this
way, the person is trained with feedback on the probabilities of faces
being assigned to categories and appropriate actions. How can this
change from learning by feedback be incoporated into the system? There
are many way to do this, but one way is to use the feedback to update
the Hamiltonians that are used to form the unitary transformations.
For example, in the category - decision making task, the unitary matrix
$U_{f}$ in the transformation $\left|\psi_{f}\right\rangle =U_{f}\left|\psi_{I}\right\rangle $
can be updated through feedback about the category. Also the unitary
matrix $U_{DC}$ that changes the basis from categorization to decision
can be updated based on feedback about the correct action.

A separate question concerns possible carry over effects from answers
on one trial to the next trial. In other words, how does the final
state at the end of one trial, say $\left|\psi_{AB},t\right\rangle $
after categorizing a face as a ``bad guy'' and deciding to ``attack'',
evolve during the intertrial interval into the initial state $\left|\psi_{I},t+1\right\rangle $
before the face is presented for the next trial? Usually the information
accumulated on one trial is not relevant for the next trial, and the
participants are instructed to treat the trials separately and independently.
The intertrial interval separating trials is made clear to the participant
and sufficiently long to prepare for the next trial (e.g. a reasonable
pause with a blank screen). The state needs to be reset during this
intertrial interval from some final state (e.g., $\left|\psi_{AB},t\right\rangle $)
after the previous trial back to a common neutral state $\left|\psi_{I},t+1\right\rangle =\left|\psi_{0}\right\rangle $
before the next trial begins. To accomplish this reset task, some
unitary operation or projection is required to change the state during
the intertrial interval. This process is an important issue but little
is understood about it. By the way, the same issue arises with more
traditional cognitive models such as Markov models, and so the problem
is not unique to quantum models.

In some experiments, however, the information accumulated during one
episode remains relevant for another episode and so both episodes
together form a trial. In this case, there is little or no intertrial
interval separating the episodes. If a person is shown a face and
asked to categorize it with respect to aggressivess, and the same
face is continued to be shown but now the person is asked to categorize
it with respect to intelligence, and the trial is defined by the pair
of episodes, then the participant can connect these two episodes together
so that the state following the answer to the first question (agressiveness)
is carried over and used as the state for answering the second question
(intelligence). In other words, no unitary transform to reset the
system intervenes between the two questions. The experimental instructions
and conditions that determine whether the reset versus the carry over
occurs is an important matter for future research \cite{Khrennikov2014}.

\subsection{Positive operator value measures}

So far we have limited our discussion to measurement defined as projectors
which satisfy $D_{A}=D_{A}^{\dagger}=D_{A}^{2}$. A more general type
of measurement is one that does not need to satisfy either of these
two properties, and instead only satisfies positivity and completeness
properties, defined below. These generalized measurements for what
are called the positive operator value measurements (POVM). As recommended
in {Khrennikov2010},  \cite{Khrennikov2014}, POVM's could be an important tool for psychologists
who need to work with more complex types of measurements. As an example,
consider the linear operator $P_{AB}=D_{A}C_{B}$, which is a measurement
of the sequence of events ``categorize as bad guy'' and then ``attack.''
If the projectors $D_{A},C_{B}$ commute, then $P_{AB}$ defines the
projector for the conjunction of these two events, which is represented
as the intersection of subspaces $A$,$B$. Of course, this implies
no order effects. If there are order effects, then the events $D_{A},C_{B}$
do not commute and the conjunction (which is commutative) is not defined.
For the non commutative case, the linear operator $P_{AB}$ is not
a projector, it does not correspond to any single subspace, and it
is not an event. Instead this measurement operator represents a sequence
of two events. There are three other measurement operators for the
other three sequences of events in this task: $P_{AG}=D_{A}C_{G},$
$P_{WB}=D_{W}C_{B},$ $P_{WG}=D_{W}C_{G}$. Note that $P_{AB}^{\dagger}P_{AB}$
is a positive operator because $\left\langle \psi|P_{AB}^{\dagger}P_{AB}|\psi\right\rangle \geq0$
for all $\left|\psi\right\rangle $, and the sum $P_{AB}^{\dagger}P_{AB}+P_{AG}^{\dagger}P_{AG}+P_{WB}^{\dagger}P_{WB}+P_{WG}^{\dagger}P_{WG}=I$
satisfies completeness. (Technically, $P_{AB}$ is the measurement
operator and $P_{AB}^{\dagger}P_{AB}$ is the positive operator corresponding
to this measurement). These two properties guarantee that the probabilities
of these four mutully exclusive and exhaustive sequences to sum to
unity. Positivity and completeness are the two requirements needed
to define a complete set of positive operator valued measurements
(POVM's). The four measurement operators provide one way to model
sequences of events when there is order dependence. A more general
way to model conjunctions of events in psychology using a POVM formulation
was recently proposed by \cite{Miyadera2012}.  Applications of POVMs to the problems
of decision making in the framework of theory of open quantum systems were considered 
in \cite{Asano}. General discussion on a possibility to describe cognitive phenomena 
solely by using only observables represented by Hermitian 
operators can found in \cite{Khrennikov2009}. Here it was shown that some statistical 
data from cognitive psychology cannot be represented with the aid of Hermitian operators;
one has to use POVMs and even their generalizations.

\subsection{Constructing mixed states}

The state vector $\left|\psi\right\rangle $ is called a pure state.
This pure state can also be expressed as a density operator $\rho=\left|\psi\right\rangle \left\langle \psi\right|$
and then the probability of an event, such as to categorize a face
as ``bad guy,'' is given by the rule $p(C_{B})=tr[C_{B}\cdot\rho]$,
which is equivalent to the previously defined rule. However, this
representation of state allows one to define a more general mixture
state $\rho=\sum p_{j}\left|\psi_{j}\right\rangle \left\langle \psi_{j}\right|$
, $0\leq p_{j}\leq1$, $\sum p_{j}=1,$ and apply the same rule for
computing probabilities from densities. The advantage of using the
density operator is that this more general form can be used to represent
a probability obtained from a mixture of participants, where each
participant is represented by a different pure state \cite{Camparo2013}.
One complicating factor that arises when working mixed states is that
their decomposition is not unique. Given a particular mixed density
operator, there is an infinite number of ways to decompose it into
pure states. This is not necessarily bad because mixed states provide
a more general way to represent uncertainty.

\section{Alternative interpretations}
\footnote{This comparison was inspired by working together on \cite{Khrennikov2014},
as well as \cite{DzhafAtman2014}}

\subsection{Quantum - cognition system}

We have presented one view of the mapping of psychological concepts
into quantum concepts. Let us briefly summarize them more formally.
First we choose the dimension $N$ of the Hilbert space. The stimulus
that is presented to the participant (e.g., stimulus $f)$ along with
other experimental information prepares the state $\left|\psi_{f}\right\rangle $.
A specific question (is the face to be presented a ``bad guy?'',
is it ``handsome,?'' ``does the person look intelligent?,'' ``should
you attack this face?,'' or whatever question you wish) determines
the basis for the Hilbert space. The basis vectors are used to define
the projectors (e.g., $D_{A},D_{W}$), and a linear combination of
the projectors forms the observable (e.g. observable $D)$. Let us
call this the ``quantum cognition'' system \cite{BusemeyerBruza2012}.
Now we consider some alternative systems that have been proposed.

\subsection{Stimulus-response system}

Another approach, called the Stimulus-Response system, does not assume
that the stimulus changes the state. Instead the state only depends
on the individual $\left|\psi_{I}\right\rangle $ (see, e.g., \cite{Khrennikov2014})
. The stimulus and question together define the observable $D(f)$
. Consider again the categorization-decision task in which the person
sees a face and then decides an action. Before any face or other stimulus
information is presented, the person is in the initial state $\left|\psi_{I}\right\rangle $.
The face stimulus together with the decision determine the observable
$D(f)$. Actually this mapping turns out not to be very different
than the quantum cognition mapping. Starting from the quantum cognition
system, we have $$
p(A|f)=\left\Vert D_{A}\cdot\left|\psi_{f}\right\rangle \right\Vert ^{2}
$$
$$
=\left\Vert D_{A}\cdot U_{f}\cdot\left|\psi_{I}\right\rangle \right\Vert ^{2}
$$
$$
\left\Vert \left(U_{f}^{\dagger}\cdot D_{A}\cdot U_{f}\right)\cdot\left|\psi_{I}\right\rangle \right\Vert ^{2}
$$
$$
=\left\Vert D_{A}(f)\cdot\left|\psi_{I}\right\rangle \right\Vert ^{2}
$$
which is the stimulus response system with $D_{A}\left(f\right)=U_{f}^{\dagger\cdot}D_{A}\cdot U_{f}$.
In this way, the stimulus-response system is a generalization of quantum
cognition system; alternatively, the quantum cognition system unpacks
and breaks the general function of the stimulus - response system
down into its cognitive components. This difference between the two
systems corresponds to the difference between the Heisenberg and Schrödinger
pictures for a quantum system.

\subsection{State Context Property system}

Aerts and Gabora and colleagues apply slightly different rules, which
they call the State Context Property or SCoP system \cite{AertsGabora2005}.
They frequently work with conceptual combination problems. For example,
a person may be informed that they will be considering the concept
of say ``pet insect.'' They are shown an example, such as a spider,
and then they are asked to decide whether or not the example is a
member of the concept ``pet insect.'' When asked to consider a concept,
like a pet, the person starts in a ``ground'' state for the pet
concept denoted $\left|\psi_{P}\right\rangle $. When asked to consider
the concept of pet in the context of it being an insect ``is this
a pet insect'', the ground state is projected onto the subspace for
this insect context to produce a new state $\left|\psi_{PI}\right\rangle $.
The experimenter then asks a question: is this example spider a member
of the category pet insect? The pet insect state is then defined as
a superposition with regard to the example, i.e., superimposed about
whether or not a spider is a member of pet insect. The yes, no answers
to the membership question correspond to the projectors $M_{Y},M_{N}=\left(I-M_{Y}\right)$,
respectively, and together they determine the membership observable
$M=1\cdot M_{y}+0\cdot M_{n}$. The same membership observable $M$
is used for all membership questions regardless of the examples that
are presented. If the example is changed from spider to beetle, then
the same observable $M$ is applied to this new question.

Now we apply the SCoP system to the categorization - decision task
following \cite{Aerts2009}. We can define the ground state $\left|\psi_{U}\right\rangle $,
as the concept ``do I want to attack this face?'' when the category
is unknown or undecided. When placed into the context ``this is a
good guy'' the state changes to $\left|\psi_{G}\right\rangle $,
but when placed into the context ``this is a bad guy'' the state
changes instead to $\left|\psi_{B}\right\rangle $. The observable
$M$ represents (yes,no), which is applied to all three of these states
to determine the probability of ``attack.'' In \cite{Aerts2009},
a simple 3 dimensional model was used to work out a SCoP model in
detail. In this simple 3-dimensional model, the states $\left|\psi_{G}\right\rangle $,$\left|\psi_{B}\right\rangle $
were designed to be orthogonal, and the unknown state $\left|\psi_{U}\right\rangle =\frac{1}{\sqrt{2}}\left(\left|\psi_{G}\right\rangle +\left|\psi_{B}\right\rangle \right)$
was assumed to be a superposition of the two known states.

SCoP sometimes works differently than the quantum cognition system
and the stimulus response system. If we applied the quantum cognition
system to the conceptual combination problem, then the person's state
$\left|\psi_{s}\right\rangle $ is prepared by the example ``spider,''
which is the stimulus that is displayed to the person The person knows
that he or she is dealing with a spider, and this is not uncertain.
What is uncertain is whether or not it is a pet, or more speficifically,
whether or not it is a pet insect. The question ``is this a pet''
is represented by one observable $M_{P}$, and the question ``is
this an insect'' is represented by another observable $M_{I}$. Likewise,
the stimulus-response system would form the observable $M(s,P)$ from
the combination of the stimulus (spider) and question (is this a pet?).

\subsection{Comparison }

Let's see how these systems work with a different example. (This section
is related to issues recently brought up by \cite{DzhafAtman2014}.)
Quantum theory has been successfully applied to question order effects
in attitude surveys \cite{WangInPress}. Suppose a person is asked
to judge whether or not a political administrator (yet to be presented)
is honest and trustworthy. For half of the respondents, the picture
and name of one administrator (e.g, Clinton) is shown first and a
judgment is made, and this followed by the picture and name of another
administrator (e.g., Gore) and another judgment is made; for the other
half of the respondants, the pictures and names are shown in the opposite
order. These questions form two related episodes, answers to one are
relevant for answering the other, and they are asked back to back
with little or no time interval between questions, and so they can
be treated as one trial. Completely unrelated types of questions are
presented on other trials.%
\footnote{(This is the actual procedure for some of the surveys examing order
effects, see \cite{Wangetal2013}).%
} For these closely related type of trials with back to back measurements,
large order effects are observed with all sorts of attitude questions.
Consider the probability of saying ``yes'' to the Clinton question
and then ``yes'' to the Gore question.

Using the stimulus-response approach, we define $C_{y}$ as the projector
for the answer ``yes Clinton is honest and trustworty'' to the Clinton
stimulus, and we define $G_{y}$ as the projector for the answer ``yes
Gore is honest and trustworty'' to the Gore stimulus. Using the stimulus
response approach we obtain the result $p(Cy,Gy)=\left\Vert G_{y}C_{y}\left|\psi_{I}\right\rangle \right\Vert ^{2}$.
Now suppose we define the projectors as follows $C_{y}=U_{CI}M_{y}U_{CI}^{\dagger}$,
$G_{y}=U_{GI}M_{y}U_{GI}^{\dagger}$ where $M_{y}$ is the generic
projector for ``yes he is honest and trustworthy'' applicable to
any person, $U_{CI}$, $U_{GI}$ are a unitary operators. The projector
$U_{CI}M_{y}U_{CI}^{\dagger}$ represents the idea that we are examining
the issue of honest and trustworthy from the Clinton perspective.
Alternatively $U_{GI}M_{y}U_{GI}^{\dagger}$ represents the same question
but now from the Gore perspective.

Now we show that the predictions from the quantum cognition system
agree with those from the stimulus - rsponse system for this example.
The stimulus-response system expresses the probability as (see Appendix
for more details) 
\begin{eqnarray*}
p(Cy,Gy) & = & \left\Vert \left(U_{GI}M_{y}U_{GI}^{\dagger}\right)\left(U_{CI}M_{y}U_{CI}^{\dagger}\right)\left|\psi_{I}\right\rangle \right\Vert ^{2}\\
 & = & \left\Vert M_{y}U_{GC}^{\dagger}M_{y}U_{CI}^{\dagger}\left|\psi_{I}\right\rangle \right\Vert ^{2}\\
 & = & \left\Vert \left(U_{GC}M_{y}U_{GC}^{\dagger}\right)M_{y}\left(U_{CI}^{\dagger}\left|\psi_{I}\right\rangle \right)\right\Vert ^{2}\\
 & = & \left\Vert \left(U_{GC}^{\dagger}M_{y}U_{GC}^{\dagger}\right)M_{y}\left|\psi_{C}\right\rangle \right\Vert ^{2},
\end{eqnarray*}
 where $U_{GC}^{\dagger}=U_{GI}^{\dagger}U_{CI}$ and $\left|\psi_{C}\right\rangle =U_{CI}^{\dagger}\left|\psi_{I}\right\rangle $
and we made use of the length preserving property of unitary operators.
The latter expression is the result obtained from the quantum cognition
approach -- the initial state (before the first stimulus) is unitarily
transformed by the Clinton stimulus, then the projector for the question
``yes he is honest and trustworthy'' is applied, and then the resulting
state is unitarily transformed by the Gore stimulus, and then the
projector for ``yes he is honest and trustworthy'' is reapplied
to the new state.

In closing, it remains an empirical question whether one system will
ultimately provide a better representation for psychological studies
as compared to another. The field is too new and we need to be in
an exploratoy mode. All three systems need to be investigated in a
more competitive way to see which one evolves and survives to become
most successful.

\section{Appendix}

The purpose of this appendix is to analyze the sequence of projections
$$
\left(U_{GN}M_{y}U_{GN}^{\dagger}\right)\left(U_{CN}M_{y}U_{CN}^{\dagger}\right)\left|\psi_{0}\right\rangle.
$$
Suppose the Hilbert space is $N$- dimensional. We will focuse on
the use of three different bases for spanning this space. One is the
``neutral'' basis $\left\{ \left|N_{i}\right\rangle ,i=1,N\right\} $
used to represent the question ``is the person honest and trustworthy''
for \emph{any} person, which is used in the quantum cognition system.
The other two $\left\{ \left|C_{i}\right\rangle ,i=1,N\right\} $$,\left\{ \left|G_{i}\right\rangle ,i=1,N\right\} $
are used to represent the ``Clinton/Gore questions about honest and
trustworthy used in the stimulus - response system.

Define $U_{CN}=\sum_{k}\left|C_{k}\right\rangle \left\langle N_{k}\right|$
as the unitary operator used to change from the neutral to the Clinton
basis; $U_{GN}=\sum_{k}\left|G_{k}\right\rangle \left\langle N_{k}\right|$
changs from the neutral to the Gore basis, and it follows that 
$$
U_{GC}=U_{GN}U_{CN}^{\dagger}
$$
$$
=\left(\sum_{j}\left|G_{j}\right\rangle \left\langle N_{j}\right|\right)\left(\sum_{j}\left|N_{j}\right\rangle \left\langle C_{j}\right|\right)
$$
$$
=\sum_{i,j}\left|G_{i}\right\rangle \left\langle N_{i}\right|\cdot\left|N_{j}\right\rangle \left\langle C_{j}\right|
$$
$$
=\sum_{k}\left|G_{k}\right\rangle \left\langle C_{k}\right|
$$
changes to from the Gore to the Clinton basis. 

When expressed in terms of either the neutral or Clinton basis, the
$N\times N$ matrix for $U_{CN}$ equals $V_{CN}=\left[\left\langle N_{j}|C_{i}\right\rangle \right],$
(transition to row $j$ from column $i$), with 
$$
\left\langle N_{j}|C_{i}\right\rangle =\left\langle N_{j}\right|\sum_{k}\left|C_{k}\right\rangle \left\langle N_{k}\right|\cdot\left|N_{i}\right\rangle
$$
$$
=\left\langle C_{j}\right|\sum_{k}\left|C_{k}\right\rangle \left\langle N_{k}\right|\cdot\left|C_{i}\right\rangle 
$$
$$
=\left\langle N_{j}|C_{i}\right\rangle.
$$ Note that $V_{CN}^{\dagger}=\left\langle C_{j}|N_{i}\right\rangle.
$
Likewise, when expressed in terms of the either the neutral or Gore
basis, the $N\times N$ matrix for $U_{GN}$ equals $V_{GN}=\left[\left\langle N_{j}|G_{i}\right\rangle \right]$,
and $V_{GN}^{\dagger}=\left\langle G_{j}|N_{i}\right\rangle .$ When
expressed in terms of the either the Clinton or Gore basis, the $N\times N$
matrix for $U_{GC}$ equals $V_{GC}=V_{CN}^{\dagger}V_{GN}=\left[\left\langle C_{j}|G_{i}\right\rangle \right],$
and note that the matrix for $U_{GC}^{\dagger}$ equals $V_{GC}^{\dagger}=V_{GN}^{\dagger}V_{CN}=\left[\left\langle G_{j}|C_{i}\right\rangle \right].$ 

Define the initial state, at the trial beginning and before any stimulus
is presented, as $\left|\psi_{0}\right\rangle =$$\sum_{i}\left|N_{i}\right\rangle \left\langle N_{i}|\psi_{0}\right\rangle $
$=$$\sum_{i}\left|C_{i}\right\rangle \left\langle C_{i}|\psi_{0}\right\rangle $
$=\sum_{i}\left|G_{i}\right\rangle \left\langle G_{i}|\psi_{0}\right\rangle $
and $\left\Vert \left|\psi_{0}\right\rangle \right\Vert ^{2}=1.$
When expressed in terms of the neutral basis, the $N\times1$ matrix
of coordinates equals $V_{N}(0)=\left[\left\langle N_{i}|\psi_{0}\right\rangle \right]$,
when expressed in terms of the Clinton basis, the $N\times1$ matrix
of coordinates equals $V_{C}(0)=\left[\left\langle C_{i}|\psi_{0}\right\rangle \right]$,
and when expressed in terms of the Gore basis, the $N\times1$ matrix
of coordinates equals $V_{G}(0)=\left[\left\langle G_{i}|\psi_{0}\right\rangle \right]$. 

Finally, we define the projector for the answer ``yes'' to the question
``is the person honest and trustworthy'' for any person. This is
defined in terms of the neutral basis as $M_{y}=\sum_{j\in yes}\left|N_{j}\right\rangle \left\langle N_{j}\right|.$
When expressed in terms of the neutral basis, this equals an $N\times N$
indicator matrix $W_{y}=\left[w_{ji}\right]$, $w_{ji}=\left\langle N_{j}|M_{y}|N_{i}\right\rangle $
with ones on the diagonal corresponding to yes, and zero otherwise. 

We start by analyzing the first the Clinton question $\left(U_{CN}M_{y}U_{CN}^{\dagger}\right)\left|\psi_{0}\right\rangle $.
First, we obtain $$
U_{CN}^{\dagger}\left|\psi_{0}\right\rangle 
$$ 
$$=
\sum_{k}\left|N_{k}\right\rangle \left\langle C_{k}\right|\cdot\sum_{j}\left|N_{j}\right\rangle \left\langle N_{j}|\psi_{0}\right\rangle 
$$
$$
=\sum_{k}\left|N_{k}\right\rangle \sum_{j}\left\langle C_{k}|N_{j}\right\rangle \left\langle N_{j}|\psi_{0}\right\rangle. 
$$
This corresponds to the matrix product $V_{CN}^{\dagger}\cdot V_{N}$.
Second, we apply the measurement to obtain $$
M_{y}U_{CN}^{\dagger}\left|\psi_{0}\right\rangle =
$$
$$
\sum_{k\in yes}\left|N_{k}\right\rangle \left\langle N_{k}\right|\cdot
\sum_{j}\left|N_{j}\right\rangle \sum_{i}\left\langle C_{j}|N_{i}\right\rangle \left\langle N_{i}|\psi_{0}\right\rangle 
$$
$$
=\sum_{k\in yes}\left|N_{k}\right\rangle \sum_{j}\left\langle C_{k}|N_{j}\right\rangle \left\langle N_{j}|\psi_{0}\right\rangle 
$$,
which corresponds to the matrix $W_{y}\cdot V_{CN}^{\dagger}\cdot V_{N}(0).$
Third, we apply the last unitary operator to obtain the projection
$\left|\psi_{1}\right\rangle =\left(U_{CN}M_{y}U_{CN}^{\dagger}\right)\left|\psi_{0}\right\rangle $$=\sum_{k}\left|C_{k}\right\rangle \left\langle N_{k}\right|\cdot\sum_{k\in yes}\left|N_{k}\right\rangle \sum_{j}\left\langle C_{k}|N_{j}\right\rangle \left\langle N_{j}|\psi_{0}\right\rangle $
$=\sum_{k\in yes}\left|C_{k}\right\rangle \sum_{j}\left\langle C_{k}|N_{j}\right\rangle \left\langle N_{j}|\psi_{0}\right\rangle $.
This corresponds to the matrix product $V_{C}(1)=\left(V_{CN}\cdot W_{y}\cdot V_{CN}^{\dagger}\right)\cdot V_{N}(0).$
Using the fact that $\left\langle C_{j}|\psi_{0}\right\rangle =\sum_{j}\left\langle C_{k}|N_{j}\right\rangle \left\langle N_{j}|\psi_{0}\right\rangle $
the result of the first measurement can be expressed as $\left|\psi_{1}\right\rangle =$$\sum_{k\in yes}\left|C_{k}\right\rangle \left\langle C_{j}|\psi_{0}\right\rangle $.
The probability of yes to the first question equals the squared length
$\left\Vert \left|\psi_{1}\right\rangle \right\Vert ^{2}=V_{C}^{\dagger}(1)V_{C}(1)\leq1.$
(Notice that we have not normalized this projection).

Hereafter, we operate on the projection $\left|\psi_{1}\right\rangle =\sum_{k\in yes}\left|C_{k}\right\rangle \left\langle C_{j}|\psi_{0}\right\rangle $
$=\sum_{k\in yes}\left|C_{k}\right\rangle \sum_{j}\left\langle C_{k}|N_{j}\right\rangle \left\langle N_{j}|\psi_{0}\right\rangle $.
This projection can be also be expressed in the Gore basis as $\left|\psi_{1}\right\rangle =\sum_{k}\left|G_{k}\right\rangle \left\langle G_{k}|\psi_{1}\right\rangle $.
It's matrix representation equals $V_{C}(1)=\left(V_{CN}\cdot W_{y}\cdot V_{CN}^{\dagger}\right)\cdot V_{N}(0).$

Now we repeat these operations for the Gore question $\left(U_{GN}M_{y}U_{GN}^{\dagger}\right)\left|\psi_{1}\right\rangle $.
First, we obtain $U_{GN}^{\dagger}\left|\psi_{1}\right\rangle $$=$$\sum_{k}\left|N_{k}\right\rangle \left\langle G_{k}\right|\cdot\sum_{j}\left|G_{j}\right\rangle \left\langle G_{j}|\psi_{1}\right\rangle $
$=\sum_{k}\left|N_{k}\right\rangle \cdot\sum_{k}\left\langle G_{k}|\psi_{1}\right\rangle $.
This corresponds to the matrix product $V_{GN}^{\dagger}\cdot V_{C}(1)=V_{GN}^{\dagger}\cdot\left(V_{CN}\cdot W_{y}\cdot V_{CN}^{\dagger}\right)\cdot V_{N}(0)$.
Then we apply the measurement to obtain $M_{y}U_{GN}^{\dagger}\left|\psi_{1}\right\rangle =\sum_{k\in yes}\left|N_{k}\right\rangle \left\langle N_{k}\right|\cdot\sum_{j}\left|N_{j}\right\rangle \left\langle G_{j}|\psi_{1}\right\rangle $
$=\sum_{k\in yes}\left|N_{k}\right\rangle \left\langle G_{k}|\psi_{1}\right\rangle $
which corresponds to the matrix $W_{y}\cdot V_{GN}^{\dagger}\cdot V_{CN}\cdot W_{y}\cdot V_{CN}^{\dagger}\cdot V_{N}(0).$
Then we apply the last unitary operator to obtain the second projecton
$\left|\psi_{2}\right\rangle =\left(U_{GN}M_{y}U_{GN}^{\dagger}\right)\left|\psi_{1}\right\rangle $$=\sum_{k}\left|G_{k}\right\rangle \left\langle N_{k}\right|\cdot\sum_{j\in yes}\left|N_{j}\right\rangle \left\langle G_{j}|\psi_{1}\right\rangle $
$=\sum_{k\in yes}\left|G_{k}\right\rangle \cdot\left\langle G_{k}|\psi_{1}\right\rangle .$
This corresponds to the matrix product $V_{G}(2)=\left(V_{GN}\cdot W_{y}\cdot V_{GN}^{\dagger}\right)\left(V_{CN}\cdot W_{y}\cdot V_{CN}^{\dagger}\right)\cdot V_{N}(0).$
The final probability of yes to the first and then the second question
equals the squared length of the second projection $\left\Vert \left|\psi_{2}\right\rangle \right\Vert ^{2}=V_{G}^{\dagger}(2)V_{G}(2)\leq\left\Vert \left|\psi_{1}\right\rangle \right\Vert ^{2}=V_{C}^{\dagger}(1)V_{C}(1)\leq1.$ 

In sum, using the convenient matrix representation, we find that the
answer from the stimulus response system equals $$
\left\Vert \left(V_{GN}\cdot W_{y}\cdot V_{GN}^{\dagger}\right)\left(V_{CN}\cdot W_{y}\cdot V_{CN}^{\dagger}\right)\cdot V_{N}(0)\right\Vert ^{2}
$$
$$
=\left\Vert W_{y}\cdot V_{GC}^{\dagger}\cdot W_{y}\cdot V_{CN}^{\dagger}\cdot V_{N}(0)\right\Vert ^{2},$$
where the effect of $V_{GN}$ in the last step is ignored because
it does not change length. The latter expression describes the same
answer in terms of the quantum cognition system: First, the initial
state $V_{N}(0)$, before any stimulus, is changed by a unitary matrix
to the state $\left(V_{CN}^{\dagger}\cdot V_{N}(0)\right)$ based
on the information provided by the stimulus. Second, a measurement
is taken using the general measurement matrix $W_{y}$ that does not
depend on any stimulus. Third the state is changed again by the unitary
matrix $V_{GC}^{\dagger}$ that depends on the second stimulus. Finally,
the second measurement is taken using the same general measurement
matrix.

\end{document}